\documentclass[a4paper,11pt]{article}
\usepackage{pos}
\usepackage{wrapfig}

\title{Towards the Composition of sub-PeV Cosmic Rays at IceCube}

\author{Julian Saffer$^{1,*}$ for the IceCube Collaboration}
\renewcommand{\printHeadAuthors}{Julian Saffer}

\affiliation[1]{Institute of Experimental Particle Physics, Karlsruhe Institute of Technology, 76021 Karlsruhe, Germany}
\affiliation[*]{$\mathrm{Speaker}$}

\emailAdd{julian.saffer@kit.edu}

\abstract{With the implementation of a low-energy trigger, the surface array of the IceCube Neutrino Observatory is able to record cosmic-ray induced air showers with a primary energy of a few hundred TeV. This extension of the energy range closes the gap between direct and indirect observations of primary cosmic rays and provides the potential to test the validity of hadronic interaction models in the sub-PeV regime. Composition analyses at IceCube highly benefit from its multi-detector design. Combining the measurement of the electromagnetic shower component and low-energy muons at the surface with the response of the in-ice array to the associated high-energy muons improves the directional reconstruction accuracy and opens unique possibilities to extract the primary particle's mass. In this contribution, a new methodical approach for the analysis of these low-energy air showers is presented, including techniques for the identification of coincident background in the in-ice detector and a machine learning model based on convolutional neural networks to determine the elemental composition. The achieved performance in primary mass discrimination and energy reconstruction of air-shower events is discussed.}

\FullConference{XVIII International Conference on Topics in Astroparticle and Underground Physics (TAUP2023)\\
28. August -- 1. September, 2023\\
University of Vienna\\}

\begin{document}
\maketitle

\section{Introduction}

\begin{wrapfigure}{r}{0.39\textwidth}
  \begin{center}
    \includegraphics[width=0.35\textwidth]{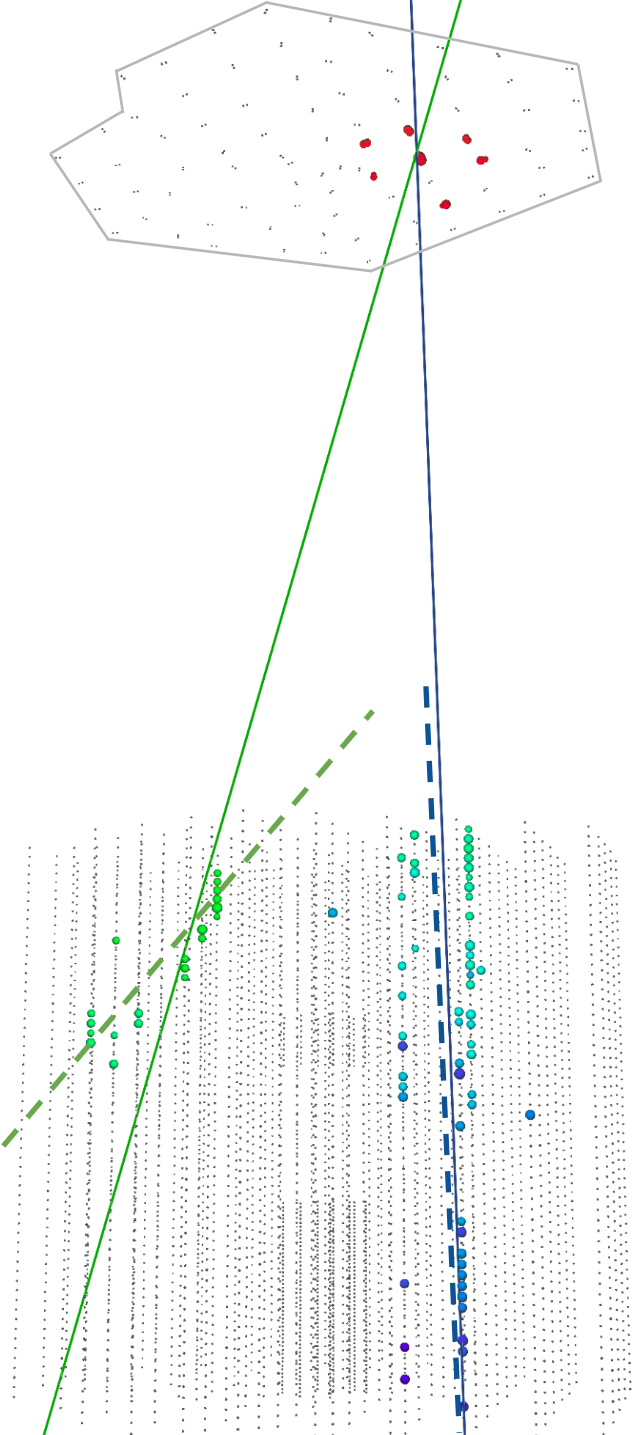}
  \end{center}
  \caption{View of a simulated vertical cosmic-ray event with background contamination seen from below the surface - IceTop on top, the in-ice array below. Bubbles represent pulses measured by DOMs. Their color is determined by the time of detection (red $\rightarrow$ blue), their size corresponds to the recorded charge. The in-ice pulses have been split into two independent pulse series (blue and green). Hypothetical tracks connecting IceTop with in-ice (solid lines) are fit to both, as well as pure in-ice fits (dashed lines). The angle $\Psi$ is measured between solid and dashed lines.}
  \label{fig:eventview}
\end{wrapfigure}

The IceCube Neutrino Observatory provides unique possibilities not only as a neutrino telescope but also as a hybrid cosmic-ray detector. Located at the geographic South Pole, IceCube consists of a cubic-kilometer optical in-ice array of 5160 digital optical modules (DOMs) and a surface array, called IceTop, comprising 81 stations, each built of two ice-Cherenkov tanks with pairs of embedded DOMs. IceTop reaches full efficiency for cosmic rays above 3\,PeV primary energy which marks the beginning of the energy range for recently published composition analyses \cite{3year, koundal2023}. By detecting both the electromagnetic component of near-vertical showers at the surface and the bundle of TeV muons deep in the ice coincidentally (Fig.~\ref{fig:eventview}), crucial information can be combined for an energy-binned composition reconstruction. The implementation of a low-energy trigger already enabled an all-particle spectrum analysis reaching below the knee to 250\,TeV \cite{lowenergy}, a region where the composition determination is increasingly uncertain for indirect cosmic-ray measurements. The first approach for reconstructing the type of cosmic-ray primaries at these low energies with IceCube has been discussed in \cite{composition2021}.

This work presents developments in the rejection of in-ice background in coincident air-shower events as well as in deep learning techniques towards a composition analysis in the sub-PeV energy range. The used Monte-Carlo (MC) simulations have been produced with Sibyll~2.3d \cite{sibyll23d} as hadronic interaction model in CORSIKA \cite{corsika} for the energy range ${5.0\leq\log_{10}\left(E_\mathrm{MC}/\mathrm{GeV}\right)<7.5}$ with proton and iron as primaries, including realistic background.

\section{Background Cleaning}

\begin{figure}
    \centering
    \begin{minipage}{0.497\textwidth}
        \centering
        \includegraphics[width=1\linewidth]{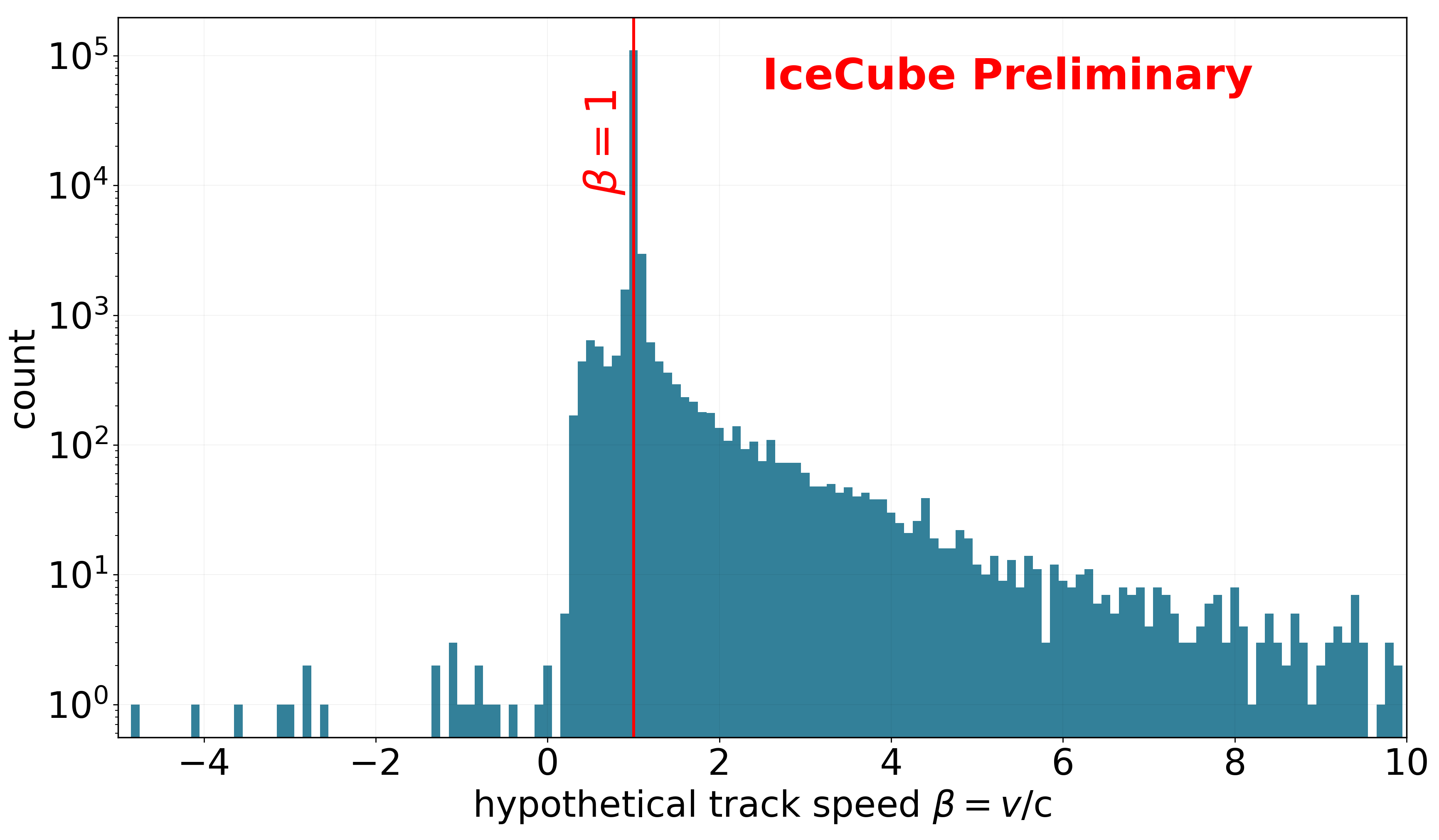}
    \end{minipage}
    \begin{minipage}{0.497\textwidth}
        \centering
        \includegraphics[width=1\linewidth]{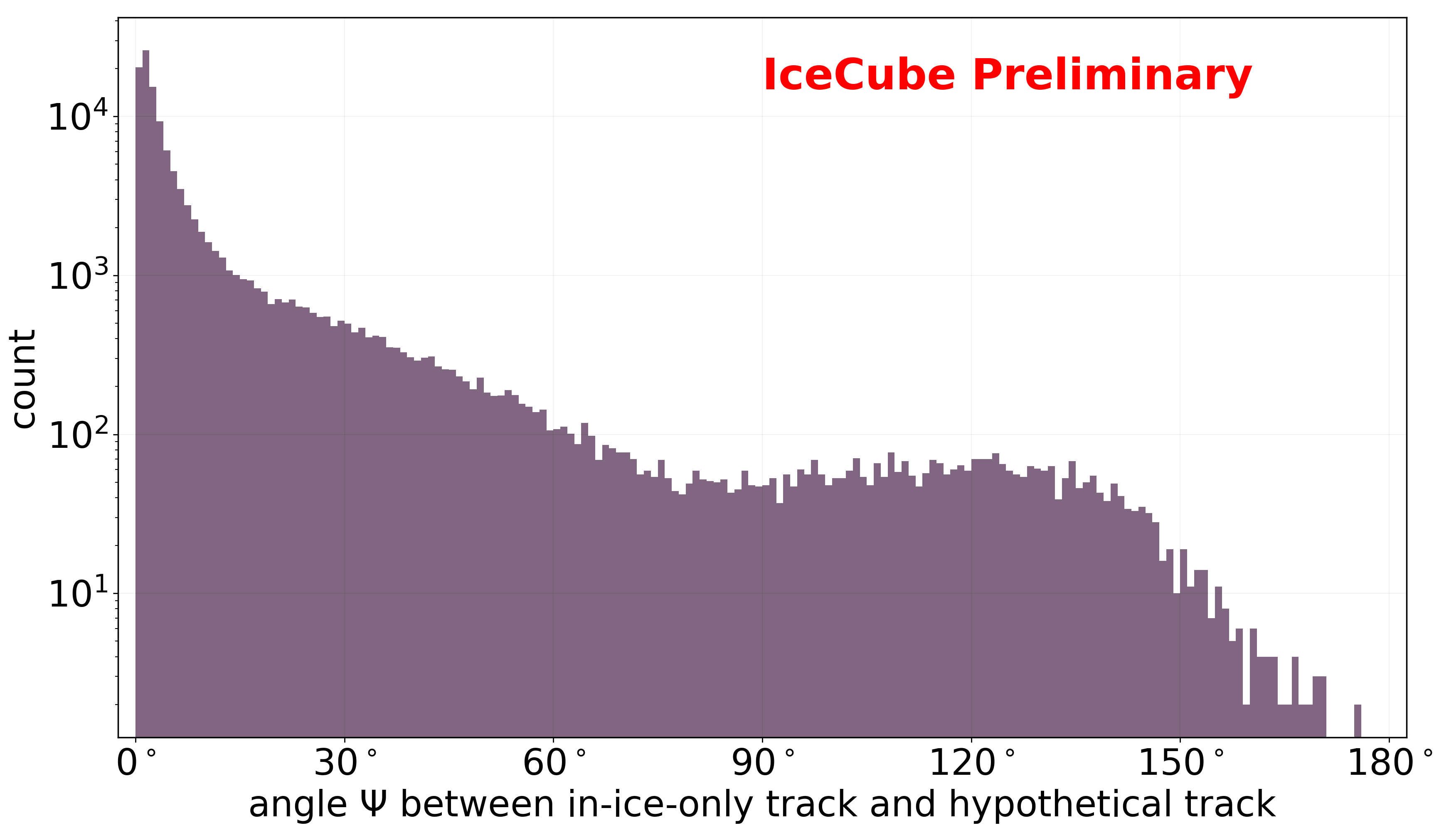}
    \end{minipage}
    \caption{Distributions of hypothetical track speed $\beta$ (left) and the angle $\Psi$ between in-ice track and hypothetical track (right) for 24 hours of data.}
    \label{fig:burnsample_distributions}
\end{figure}

When selecting coincident events, i.e. air showers that are detected by both IceTop and in-ice, it is inevitable that occasionally unrelated muons cross the IceCube array within the readout time window. The background cleaning used in \cite{3year, koundal2023} relies on a converging track fit to the IceTop pulses alone which is generally not available below PeV energies. By splitting the in-ice pulses in such cases into independent pulse series (blue and green in Fig.~\ref{fig:eventview}), signal and background tracks can be identified in two ways:

When connecting the IceTop shower core with the center of one in-ice pulse series both in space and time, the speed of such a hypothetical track is expected to be close to the speed of light for true signal coincidences while this does not hold for unrelated background coincidences. The left panel in Fig.~\ref{fig:burnsample_distributions} shows the distribution of the track speed which peaks at $\beta=1$ emerging from coincident events detected by both IceTop and in-ice. Deviating values correspond to background coincidences which would require a nonphysical track speed under the assumption that the muons are part of the air-shower event.

In order to complement this speed-based selection, the angle between the hypothetical shower axis and an in-ice-only track fit is taken into account. While the in-ice-only track is expected to point back to the signal deposit at the surface, the direction of background tracks is not correlated with the connection to IceTop. The right plot in Fig.~\ref{fig:burnsample_distributions} shows a peak at small angles $\Psi$ (signal) and a tail towards larger values (background). Cuts on both $\beta$ and $\Psi$ can be applied to remove pulse series that were primarily caused by background muons.

\section{Input and Architecture of CNN}

\begin{figure}
    \centering
    \includegraphics[width=1\linewidth]{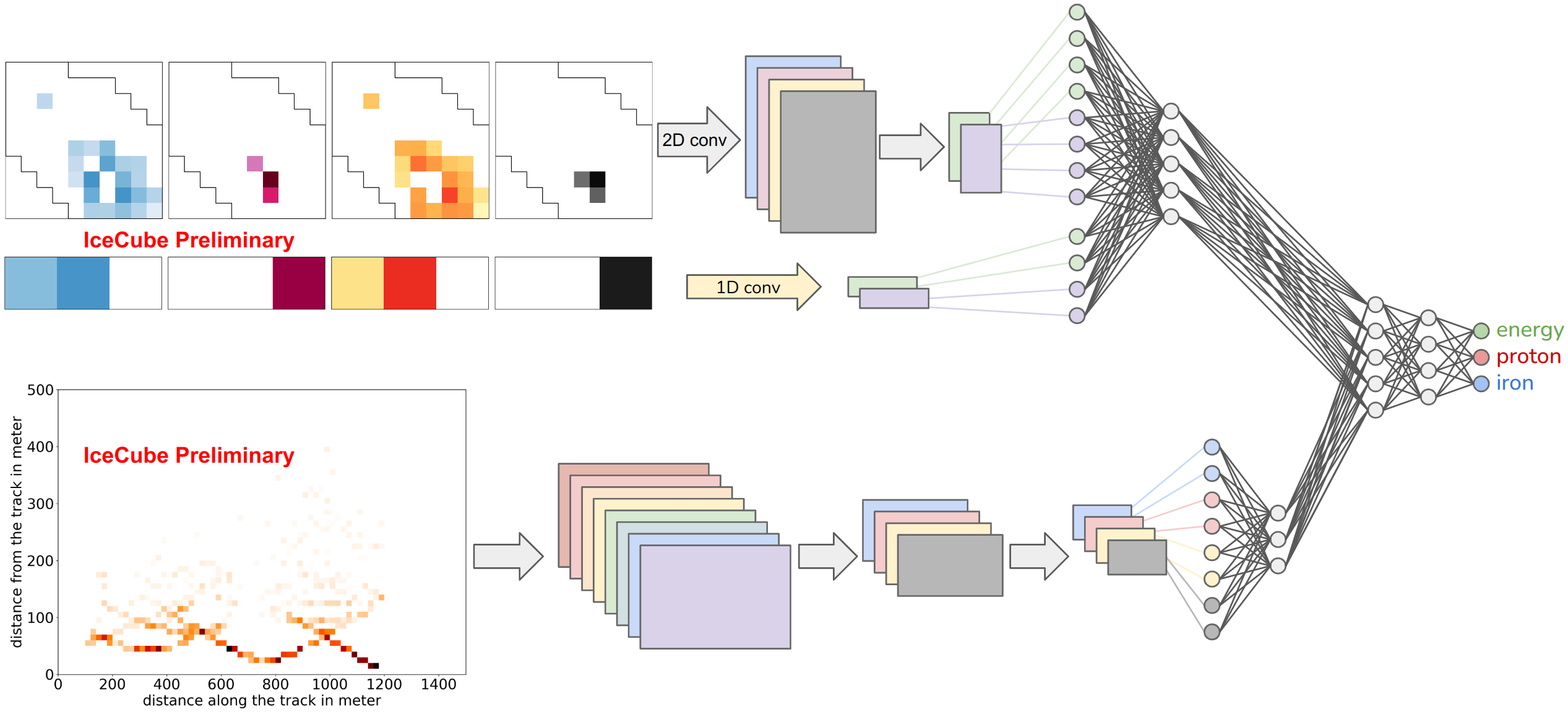}
    \caption{Architecture of the CNN. The top left block shows an example of IceTop input, the $10\times10$ images on top, the 1-dimensional arrays representing the three in-fill stations below. The darkness of each pixel visualizes the charge recorded by the respective DOM. The four image layers are shown in different color shades. The bottom left block gives an example of a 2D-histogram of in-ice pulses with slant depth (20\,m binning) on the horizontal and distance from the track (10\,m binning) on the vertical axis. All three parts of the network input are fed into separate convolution blocks that are depicted by arrows. After flattening of the intermediate outputs (colored nodes), the two IceTop blocks and the in-ice part get combined in a fully-connected network. The number of hidden nodes is reduced for better visibility. The CNN is trained for the three output nodes to predict the primary energy and to give a proton and iron score - whichever is higher determines the prediction.}
    \label{fig:network}
\end{figure}

Convolutional neural networks (CNNs) have a wide range of applications like image recognition but also include IceCube analyses that aim to distinguish between neutrino flavors \cite{yu2023}. In a similar fashion, detected charge distributions from IceTop and the in-ice array are fed into a CNN. The targets of this network are the regression of primary energy as well as the classification of the primary type of air showers. IceTop signals are brought into the shape of $4\times10\times10$ tensors - 4 being the number of DOMs per IceTop station which are interpreted as image layers - while the three in-fill stations are represented by $4\times3$ tensors (top left section of Fig.~\ref{fig:network}), both in analogy to \cite{abbasi2021convolutional}. In-ice signals are binned in two dimensions, slant depth along and distance from the reconstructed track that was fit to IceTop and in-ice signal simultaneously (bottom left section in Fig.~\ref{fig:network}). This dimension reduction from the volumetric detector array to a flat ``image'' is possible because of the radial symmetry of vertical air showers. For each event, this 2D-histogram and both surface inputs (IceTop and in-fill) are fed as tensors into respective CNN blocks which get connected to a single network with the primary energy as one target and one additional output node per primary type.

\section{Energy and Primary Type Reconstruction}

\begin{figure}[h]
    \centering
    \begin{minipage}{0.497\textwidth}
        \centering
        \includegraphics[width=1\linewidth]{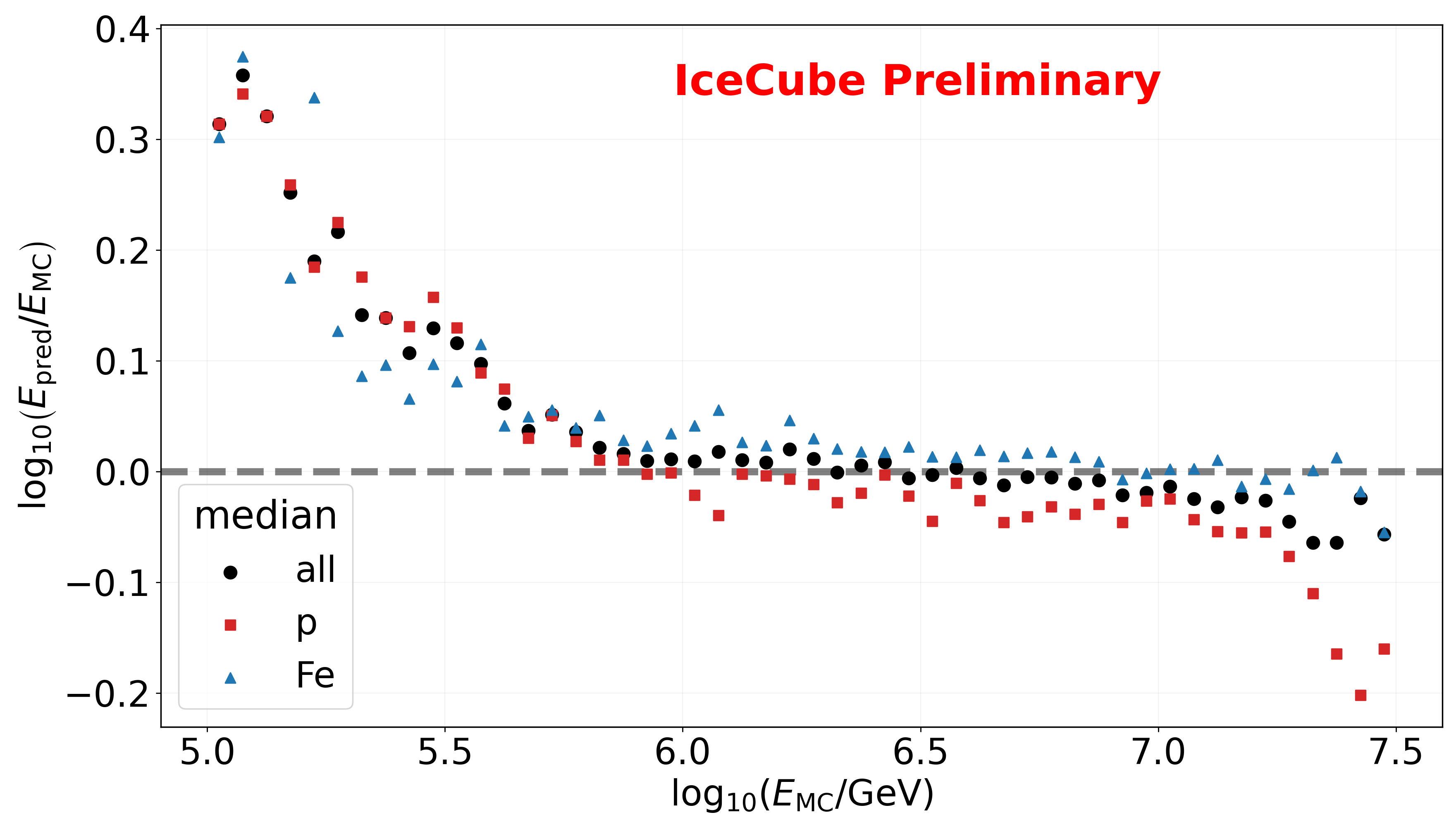}
    \end{minipage}
    \begin{minipage}{0.497\textwidth}
        \centering
        \includegraphics[width=1\linewidth]{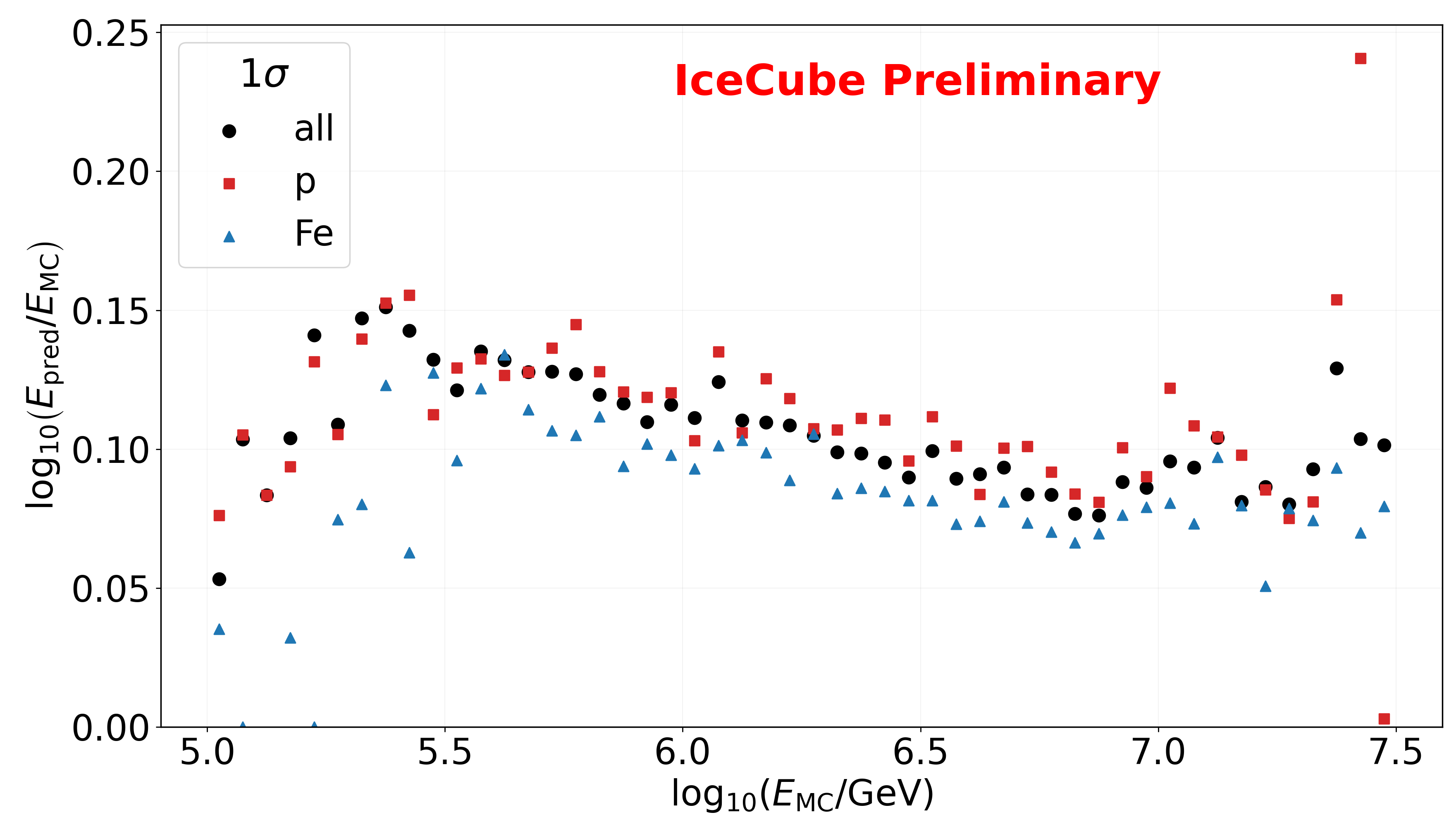}
    \end{minipage}
    \caption{Bias (left, median) and resolution (right, 68-percentile after bias subtraction) of the energy predicted by the CNN. Black points show the entire test dataset, red and blue denote proton and iron primaries, respectively.}
    \label{fig:energy}
\end{figure}

The network's energy output is directly evaluated as shown in Fig.~\ref{fig:energy}. The median of $\log_{10}\left(E_\mathrm{pred}/E_\mathrm{MC}\right)$ is chosen as bias which gets subtracted before calculating the 68-percentile of the absolute values to determine the resolution. The energy resolution $\sigma_E=\sigma\left(E_\mathrm{pred}/E_\mathrm{MC}\right)-1$ ranges from $\sim$42\% for showers at around 250\,TeV to $\sim$19\% at 7\,PeV primary energy.

\begin{figure}
    \centering
    \includegraphics[width=1\linewidth]{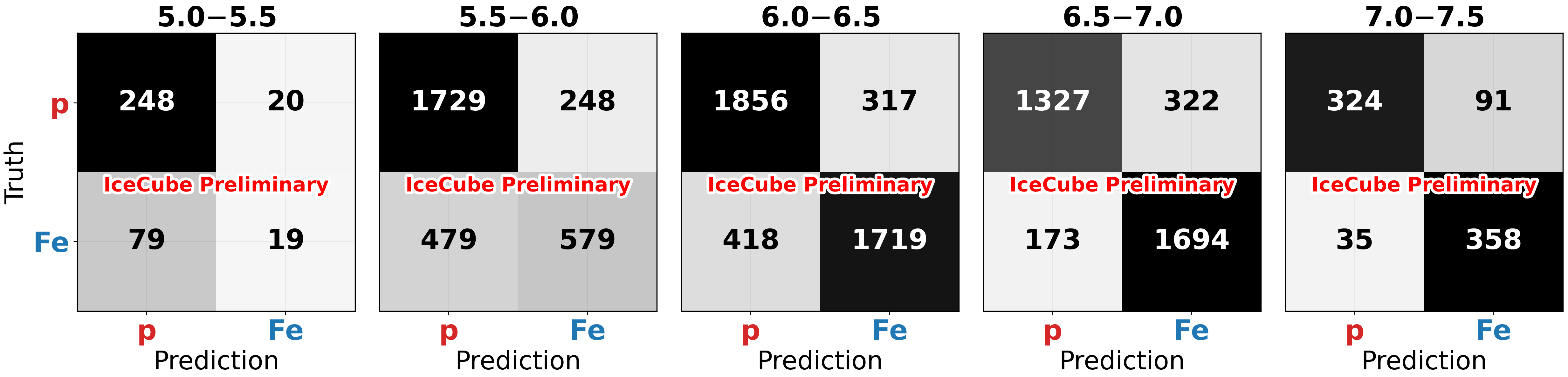}
    \caption{Confusion matrices for the binary classification of proton and iron. The range in $\log_{10}\left(E_\mathrm{MC}/\mathrm{GeV}\right)$ is indicated above. Low statistics at both ends of the energy range and an imbalance in the lowest energies are evident.}
    \label{fig:type}
\end{figure}

For the classification of the primary type, the dedicated output nodes are compared. The node with the highest value determines the prediction. Since the dataset contains proton and iron primaries, this turns into a binary classification. The confusion matrices for 5 energy ranges are presented in Fig.~\ref{fig:type}. In the sub-PeV range an accuracy of $\left(75.7^{\,+0.8}_{-0.8}\right)\%$ is reached while $\left(84.3^{\,+0.4}_{-0.5}\right)\%$ are achieved for primary energies above 1\,PeV.

\section{Discussion and Outlook}

A new approach for the rejection of background pulses in coincident cosmic-ray events at IceCube has been presented. Previous coincidence analyses require an IceTop-only reconstruction for background cleaning. The new method uses the speed of and the angle with a hypothetical track connecting IceTop with the in-ice detector which makes analyses below PeV primary energies possible. The cleaned events are then fed into a new CNN model which not only makes use of transversal symmetries but also rotational symmetry of the in-ice signal and predicts primary energy as well as type.

The energy bias shows the common behavior of over-predicting at the lowest energies and under-predicting at the highest energies. Similarly, the resolution suffers at the edges of the energy range which also is due to the lack of statistics in those energy bins as shown in Fig.~\ref{fig:type}. The energy resolution shows improvement over \cite{lowenergy} by a factor $\sigma_E^\mathrm{this\ work}/\sigma_E^{\scriptsize\cite{lowenergy}}$ of about 0.5 for $E_\mathrm{MC}$ below 1\,PeV and of approximately 0.7 up to 4\,PeV. Even though a bias known from MC can be corrected, the resolution clearly shows a composition dependency over most of the energy range as it has been observed before in \cite{3year}.

The combination of an imbalanced simulation dataset and a composition-dependent trigger efficiency of IceCube for sub-PeV cosmic rays leads to a biased classifier in the lowest energy bins. Common metrics like accuracy and area under the receiver operating characteristic curve improve with increasing primary energy.

Using complete simulation datasets of the entire energy range and balancing the training set is expected to reduce those biases. Moreover, simulations with intermediate primary masses and different hadronic interaction models are being prepared.

\vfill
\noindent \footnotesize \textbf{Acknowledgement:} This work was performed on the HoreKa supercomputer funded by the Ministry of Science, Research and the Arts Baden-Württemberg and by the Federal Ministry of Education and Research of Germany.

\normalsize
\bibliographystyle{ICRC}
\bibliography{refs}

\end{document}